\documentclass[usenatbib]{mnras}

\usepackage{newtxtext,newtxmath}

\usepackage[T1]{fontenc}
\usepackage{times,ae,aecompl}

\usepackage{graphicx}	
\usepackage{amsmath}	
\usepackage{amssymb}	
\usepackage{wasysym}
\usepackage{xcolor}
\usepackage{multirow}
\usepackage{pdfcolfoot}
\usepackage{float}
\usepackage{bigfoot}

\usepackage{etoolbox}
\makeatletter
\patchcmd\@combinedblfloats{\box\@outputbox}{\unvbox\@outputbox}{}{\errmessage{\noexpand patch failed}}
\makeatother

\newcommand{\be}{\begin{eqnarray}}
\newcommand{\ee}{\end{eqnarray}}
\newcommand{\hMpc}{h \, {\rm Mpc}^{-1}}
\newcommand{\Mpch}{h^{-1} {\rm Mpc}}

\definecolor{darkgreen}{cmyk}{0.85,0.2,1.00,0.05}

\title[Non-linear reaction to massive neutrinos]{On the road to percent accuracy III: non-linear reaction of the matter power spectrum to massive neutrinos}

\author[Cataneo et al.]{
M. Cataneo$^{1}$\thanks{E-mail: matteo@roe.ac.uk}, 
J.D. Emberson$^{2}$,
D. Inman$^{3}$,
J. Harnois-D\'eraps$^{1}$,
C. Heymans$^{1,4}$
\\
$^{1}$Institute for Astronomy, University of Edinburgh, Royal Observatory, Blackford Hill, Edinburgh, EH9 3HJ, U.K.\\
$^{2}$CPS Division, Argonne National Laboratory, Lemont, IL 60439, USA\\
$^{3}$Center for Cosmology and Particle Physics, Department of Physics, New York University, 726 Broadway, New York, \\ NY, 10003, USA\\
$^{4}$German Centre for Cosmological Lensing (GCCL), Astronomisches Institut, Ruhr-Universit\"at Bochum, Universit\"atsstr.
150, \\ 44801 Bochum, Germany.\\
}

\date{Accepted XXX. Received YYY; in original form ZZZ}

\pubyear{2019}

\begin{document}
\label{firstpage}
\pagerange{\pageref{firstpage}--\pageref{lastpage}}
\maketitle

\begin{abstract}
We analytically model the non-linear effects induced by massive neutrinos on the total matter power spectrum using the \emph{halo model reaction framework} of Cataneo et al. 2019. In this approach the halo model is used to determine the relative change to the matter power spectrum caused by new physics beyond the concordance cosmology. Using standard fitting functions for the halo abundance and the halo mass-concentration relation, the total matter power spectrum in the presence of massive neutrinos is predicted to percent-level accuracy, out to $k=10 \, \hMpc$. We find that refining the prescriptions for the halo properties using $N$-body simulations improves the recovered accuracy to better than 1\%. This paper serves as another demonstration for how the \emph{halo model reaction framework}, in combination with a single suite of standard $\Lambda$CDM simulations, can recover percent-level accurate predictions for beyond-$\Lambda$CDM matter power spectra, well into the non-linear regime.
\end{abstract}




\section{Introduction}
In the standard model of particle physics, neutrinos are treated as elementary massless particles. However, it has been conclusively shown that neutrino flavour (i.e. electronic, muonic, tauonic) can change with time~\citep{Fukuda:1998,Ahmed:2004}, a phenomenon know as \emph{flavour oscillations}. For this to be possible, at least two neutrinos must possess a non-zero mass, therefore pointing to physics beyond the standard model. Since oscillation experiments measure the mass-squared splittings between the three mass eigenstates, they can only provide a lower bound on the absolute mass scale, and hence alone cannot determine the neutrino mass hierarchy~\citep{Qian:2015}.

On the other hand, the presence of massive neutrinos has profound implications for the formation and evolution of structures in the Universe~\citep{Lesgourgues:2006}. At early times, in particular at recombination, neutrinos are ultra-relativistic and so their masses do not affect the primary CMB.  At redshifts of $\sim200(m_\nu/0.1 \, \mathrm{eV})$ neutrinos become non-relativistic; however, their still large thermal velocities prevent them from clustering strongly producing a characteristic modification to the matter power spectrum.  Large-scale structure observables are therefore sensitive to the sum of neutrino masses~\citep{Marulli:2011}, with measurable effects, for instance, on the abundance of massive galaxy clusters~\citep[e.g.][]{Costanzi:2013} and two-point shear statistics~\citep[e.g.,][]{Liu:2018}. 

Upcoming wide-field galaxy surveys will map the large-scale structure of the Universe to an unprecedented volume and accuracy~\citep{Laureijs:2011,LSST:2012,Green:2012,Levi:2013}, thus challenging our ability to predict cosmological summary statistics with the required small uncertainties over the entire range of relevant scales. In particular, percent-level knowledge of the matter power spectrum in the non-linear regime is necessary to take full advantage of future cosmic shear measurements~\citep{Taylor:2018}. At present, however, all known \mbox{(semi-)analytical} methods incorporating the non-linear effects of massive neutrinos on the matter power spectrum lack sufficient accuracy to be employed in future cosmological analyses aimed at stringent and unbiased constraints of the absolute mass scale~\citep{Bird:2012,Blas:2014,Mead:2016,Lawrence:2017}.

In this paper we demonstrate how the halo model reaction framework of~\citet{Cataneo:2019} can predict the non-linear total matter power spectrum in the presence of massive neutrinos to the accuracy requirements imposed by the next generation of cosmological surveys. Sec.~\ref{sec:methods} describes our approach and the cosmological simulations used for its validation. Sec.~\ref{sec:results} presents our results, and in Sec.~\ref{sec:discussion} we discuss their implications and future applications. 

Our baseline flat $\Lambda$CDM cosmology has total matter density $\Omega_{\rm m} = 0.2905$, baryon density $\Omega_{\rm b} = 0.0473$, reduced Hubble constant $h=0.6898$, scalar spectral index $n_{\rm s} = 0.969$ and amplitude of scalar fluctuations $A_{\rm s} = 2.422 \times 10^{-9}$ at the pivot scale $k_0 = 0.002 \, {\rm Mpc}^{-1}$. In massive neutrino cosmologies we fix all parameters to their baseline values, and vary the cold dark matter (CDM) density as $\Omega_{\rm c}=\Omega_{\rm m}-\Omega_{\rm b}-\Omega_\nu$, with $\Omega_\nu$ denoting the neutrino density. For our linear calculations we use the Boltzmann code {\sc camb}\footnote{\url{https://camb.info}}~\citep{Lewis:2000}.


\section{Methods}\label{sec:methods}

\subsection{Halo model reactions with massive neutrinos}\label{sec:reactions}

The implementation of massive neutrinos in the halo model (HM) has been previously studied in~\citet{Abazajian:2005} and~\citet{Massara:2014}, with the latter finding inaccuracies as large as 20-30\% in the predicted total matter non-linear power spectrum when compared to $N$-body simulations. To reduce these discrepancies down to a few percent, \citet{Massara:2014} proposed the use of massive-to-massless neutrino HM power spectrum ratios. Here, we follow a similar strategy by extending the recently developed \emph{halo model reaction} framework~\citep[][also see~\citet{Mead:2017} for its first applications]{Cataneo:2019} to include the effect of massive neutrinos. As we shall see in Sec.~\ref{sec:results}, this approach improves the halo model performance by more than one order of magnitude, therefore reaching the target accuracy set by the next generation of galaxy surveys, albeit neglecting baryonic feedback~\citep{Chisari:2019}.

The total matter power spectrum in the presence of massive neutrinos is given by the weighted sum 
\begin{align}\label{eq:totPk}
P^{\mathrm{(m)}}(k) = (1-f_\nu)^{2} P^{\mathrm{(cb)}}(k)+2 f_\nu(1-f_\nu) P^{(\mathrm{cb}\nu)}(k)+f_\nu^{2} P^{(\nu)}(k) \, ,    
\end{align}
where $f_\nu = \Omega_\nu/\Omega_{\mathrm{m}}$, $P^{\mathrm{(cb)}}$ is the auto power spectrum of CDM+baryons\footnote{In this work we treat baryons as cold dark matter, and only account for their early-time non-gravitational interaction through the baryon acoustic oscillations imprinted on the linear power spectrum~\citep[cf.][]{McCarthy:2018}. }, $P^{(\nu)}$ is the neutrino auto power spectrum, and $P^{(\mathrm{cb}\nu)}$ is the cross power spectrum of the neutrinos and the two other matter components\footnote{In general, we drop the dependence on redshift of the power spectrum and related quantities, unless required to avoid confusion.}. 
In our halo model predictions we approximate neutrino clustering as purely linear, allowing us to replace the neutrino non-linear auto power spectrum with its linear counterpart, $P_{\mathrm{L}}^{(\nu)}$, and thus rewrite the cross power spectrum as\footnote{This approximation is motivated by the two following arguments: (i) the cross-correlation coefficient between the neutrino and CDM fields is large on all relevant scales~\citep{Inman:2015}; and (ii) although using the linear neutrino power spectrum introduces substantial errors in the cross power spectrum on small scales~\citep{Massara:2014}, due to $P^{(\nu)} \ll P^{(\mathrm{cb})}$ and the suppression factor $2 f_\nu(1-f_\nu)$ preceding $P^{(\mathrm{cb}\nu)}$ in Eq.~\eqref{eq:totPk}, the overall impact on the total matter power spectrum becomes negligible.}~\citep{Agarwal:2011,Ali-Haimoud:2013}
\be
P_{\mathrm{HM}}^{(\mathrm{cb}\nu)}(k) \approx \sqrt{P_{\mathrm{HM}}^{\mathrm{(cb)}}(k) P_{\mathrm{L}}^{(\nu)}(k)} \, .
\ee
The CDM+baryons auto power spectrum is then divided into two-halo and one-halo contributions~\citep[see, e.g.,][]{Cooray:2002},
\be\label{eq:P_HM_cb}
P_{\mathrm{HM}}^{\mathrm{(cb)}}(k) =  P_{\mathrm{L}}^{\mathrm{(cb)}}(k) + P_{1 \mathrm{h}}^{\mathrm{(cb)}}(k) \, ,
\ee
where we neglect the two-halo integral pre-factor involving the linear halo bias~\cite[see][for details]{Cataneo:2019}\footnote{This integral introduces corrections $\gtrsim 1\%$ to the two-halo term only for $k \gtrsim 0.5 \, \hMpc$\citep[see, e.g.,][]{Massara:2014}. On these scales, however, the leading contribution to the power spectrum comes from the one-halo term instead. Moreover, in this work we take the ratio of halo-model predictions, and our findings presented in Sec.~\ref{sec:fits} suggest that ignoring the two-halo correction can introduce errors no larger than 0.3\%.}. 

In the \emph{reaction} approach described in \citet{Cataneo:2019}, we must now define a \emph{pseudo} massive neutrino cosmology, which is a flat and massless neutrino $\Lambda$CDM cosmology whose linear power spectrum is identical to the total linear matter power spectrum of the \emph{real} massive neutrino cosmology at a chosen final redshift, $z_{\mathrm{f}}$, that is
\be\label{eq:P_L_pseudo}
P_{\mathrm{L}}^{\mathrm{pseudo}}\left(k, z_{\mathrm{f}}\right)=P_{\mathrm{L}}^{\mathrm{(m)}}\left(k, z_{\mathrm{f}}\right) \, .
\ee
Owing to the different linear growth in the two cosmologies, $P_{\mathrm{L}}^{\mathrm{pseudo}}$ and $P_{\mathrm{L}}^{\mathrm{(m)}}$ can differ substantially for $z > z_{\mathrm{f}}$. In the halo model language, the ratio of the \emph{real} to \emph{pseudo} non-linear total matter power spectra, i.e. the \emph{reaction}, takes the form 
\begin{align}\label{eq:react}
\mathcal{R}(k)=\frac{(1-f_\nu)^{2} P_{\mathrm{HM}}^{\mathrm{(cb)}}(k)+2 f_\nu(1-f_\nu) P_{\mathrm{HM}}^{(\mathrm{cb}\nu)}(k)+f_\nu^{2} P_{\mathrm{L}}^{(\nu)}(k)}{P_{\mathrm{HM}}^{\mathrm{pseudo}}(k)} \, ,
\end{align}
with
\begin{align}
P_{\mathrm{HM}}^{\mathrm{pseudo}}(k) = P_{\mathrm{L}}^{\mathrm{(m)}}(k)+P_{\mathrm{1h} }^{\mathrm{pseudo}}(k) \, .
\end{align}
For a mass-dependent and spherically symmetric halo profile with Fourier transform $u(k, M)$, the one-halo term is given by the integral
\be\label{eq:P1h}
P_{1 \mathrm{h}}(k)=\int \mathrm{d} \ln M \, n(M) \left(\frac{M}{\bar{\rho}}\right)^{2}\left|u\left(k, M \right)\right|^{2} \, ,
\ee
where 
\be\label{eq:hmf}
n(M) \equiv \frac{\mathrm{d} n}{\mathrm{d} \ln M}=\frac{\bar{\rho}}{M} \left[ \nu f(\nu) \right] \frac{\mathrm{d} \ln \nu}{\mathrm{d} \ln M}
\ee
is the virial halo mass function, and we use the Sheth-Tormen multiplicity function~\citep{Sheth:2002}
\be\label{eq:ST}
\nu f(\nu)=A \sqrt{\frac{2}{\pi} q \nu^{2}}\left[1+\left(q \nu^{2}\right)^{-p}\right] \exp \left[-q \nu^{2} / 2\right] \, ,
\ee
with $\{A,q,p\} = \{0.3292,0.7665,0.2488\}$~\citep{Despali:2016}. In Eqs.~\eqref{eq:hmf} and~\eqref{eq:ST} the peak height $\nu(M,z) \equiv \delta_{\mathrm{coll}}(z)/\sigma(M,z)$, where $\delta_{\mathrm{coll}}$ is the redshift-dependent spherical collapse threshold, and 
\be
\sigma^{2}(R, z)=\int \frac{\mathrm{d}^{3} k}{(2 \pi)^{3}}|\tilde{W}(k R)|^{2} P_{\mathrm{L}}(k, z) \, .
\ee
Here, $R = (3M/4\pi \bar\rho)^{1/3}$, and $\tilde{W}$ denotes the Fourier transform of the top-hat filter.

For the halo profile in Eq.~\eqref{eq:P1h} we assume the Navarro-Frank-White (NFW) profile~\citep{Navarro:1996} truncated at its virial radius $R_{\mathrm{vir}} = (3M/4\pi \bar\rho\Delta_{\mathrm{vir}})^{1/3}$, where $\Delta_{\mathrm{vir}}$ is the redshift- and cosmology-dependent virial spherical overdensity~\citep[see, e.g.,][]{Cataneo:2019}. 
In our NFW profiles calculations, we approximate the relation between the halo virial concentration and mass with the power law
\be\label{eq:Bullock}
c(M, z)=\frac{c_{0}}{1+z}\left[\frac{M}{M_{*}(z)}\right]^{-\alpha} \, ,
\ee
where the characteristic mass, $M_{*}$, satisfies $\nu(M_{*},z) = 1$, and we set the $c$-$M$ relation parameters to their standard values $c_0 = 9$ and $\alpha = 0.13$~\citep{Bullock:2001}.

For the evaluation of the one-halo term (Eq.~\ref{eq:P1h}) we use different comoving background matter densities, linear matter power spectra, and spherical collapse evolution in the \emph{real} and \emph{pseudo} cosmologies. More specifically, for the CDM+baryons component in the \emph{real} cosmology we have
\be
\bar\rho &\rightarrow& \bar\rho_{\mathrm{cb}} \, , \\
P_{\mathrm{L}} &\rightarrow& P_{\mathrm{L}}^{\mathrm{(cb)}} \, .
\ee
Then the equation of motion for the spherical collapse overdensity~\citep[see, e.g.,][]{Cataneo:2019} is independent of mass and sourced only by the CDM+baryons Newtonian potential~\citep[cf.][]{LoVerde:2014}; the flat $\Lambda$CDM background expansion is controlled by $\Omega_{\mathrm{m}}$. On the other hand, for the \emph{pseudo} cosmology 
\be
\bar\rho &\rightarrow& \bar\rho_{\mathrm{m}} \, , \\
P_{\mathrm{L}} &\rightarrow& P_{\mathrm{L}}^{\mathrm{(m)}} \, ,
\ee
while the spherical collapse dynamics is still governed by the standard $\Lambda$CDM equation with $\Omega_{\mathrm{cb}}^{\rm pseudo} = \Omega_{\mathrm{m}}^{\rm real}$.

Finally, assuming we can accurately compute the non-linear matter power spectrum of the \emph{pseudo} cosmology with methods other than the halo model~\citep[see, e.g.,][]{Giblin:2019}, the total matter power spectrum of the \emph{real} cosmology, Eq.~\eqref{eq:totPk}, can be rewritten in the \emph{halo model reaction} framework as
\be\label{eq:tot_matter_pred}
P^{\mathrm{(m)}}(k,z)=\mathcal{R}(k,z) \times P^{\rm{pseudo}}(k,z) \, .
\ee
In this work we generally use the \emph{pseudo} matter power spectrum measured from the simulations described in the next Section. However, to test the robustness of the \emph{reaction} approach to alternative $N$-body codes implementing massive neutrinos, in Sec.~\ref{sec:halofit} we employ~\cite{Bird:2012} and~\cite{Takahashi:2012} fitting formulas as proxy for the \emph{real} and \emph{pseudo} massive neutrino simulations, respectively.


\subsection{$N$-body simulations}\label{sec:sims}
We compute our fiducial non-linear power spectra and halo properties with the publicly available $N$-body code {\sc cubep$^3$m} \citep{HarnoisDeraps:2013}, which has been modified to include neutrinos as a separate set of particles \citep{Inman:2015,Emberson:2017}. We run a suite of simulations both with and without neutrino particles. In the standard massless neutrino case, particles are initialized from the Zel'dovich displacement field  \citep{Zeldovich:1970}, obtained from the combined  baryons + CDM  transfer functions, linearly evolved from $z=0$ to $z=100$. However, for the \emph{pseudo} cosmologies we generate the initial conditions from the total linear matter power spectrum of the corresponding \emph{real} massive neutrino cosmologies at $z_{\mathrm{f}} = 0$ or 1 (see Eq.~\ref{eq:P_L_pseudo}), rescaled to the initial redshift $z = 100$ with the $\Lambda$CDM linear growth function using $\Omega_{\rm m}^{\rm real}$.

In the massive neutrino case the simulations run in two phases, as unphysical dynamics sourced by the large thermal velocities (such as unaccounted for relativistic effects or large Poisson fluctuations) can occur if neutrinos are included at too high redshift\footnote{The particle initialization and the execution pipelines were improved since \citet{Inman:2015}, which is why we provide more details here \citep[see][for additional descriptions]{Inman:2017b}.}~\citep{Inman:2015}. In the first, from $z=100$ to $z=10$, only CDM particles are evolved; the neutrinos are treated as a perfectly homogeneous background component. We account for their impact on the growth factor by multiplying a $z=10$ CDM transfer function with the neutrino correction, $D(z=100)/D(z=10)$, where  $D(a)\propto a^{1-3f_\nu/5}$ \citep{Bond:1980}. The Zel'dovich displacement is also modified to account for neutrino masses, with every velocity component being multiplied by $1-3f_\nu/5$. Finally, the mass of every particle is multiplied by $1-f_{\nu}$. With this strategy, CDM perturbations are correct at $z=10$ even though we do not evolve neutrino perturbations before then. In the second phase, neutrinos are added into the code as a separate $N$-body species. For their initialization, neutrino density and velocity fields are computed at $z=10$ from {\sc camb} transfer functions, and the Zel'dovich approximation is again used to compute particle displacements and velocities. A random thermal contribution, drawn from the Fermi-Dirac distribution, is also added to their velocities. {\sc cubep$3$m} then co-evolves neutrinos and dark matter with masses weighted by $f_\nu$ and $1-f_\nu$ respectively.

In all neutrino runs, we assume a single massive neutrino contributing $\Omega_\nu h^2=m_\nu/93.14 \, {\rm eV}$~\citep{Mangano:2005}, and consider cosmologies with $m_\nu = 0.05, 0.1, 0.2, 0.4$ eV. We perform runs with $N_\nu = 3072^3$ neutrino particles and  box sizes $L_{\rm box} = 500 \,  \Mpch$ for all values of $m_\nu$ considered, as well as one set of large-volume runs with $L_{\rm box} = 1000 \,  \Mpch$ and $m_\nu = 0.4$ eV. We use $N_{\rm cb} = 1536^3$ CDM particles in the smaller boxes and $N_{\rm cb} = 3072^3$ particles in the larger boxes, corresponding to a common mass resolution of $m_{\mathrm{cb}} = 2.78 \times 10^9 \, h^{-1}M_\odot$ for the baseline $\Lambda$CDM cosmology. A common gravitational softening length of $24 \, h^{-1}{\rm kpc}$ is also used. 

\begin{figure*}
\begin{center}
\includegraphics[width=\columnwidth]{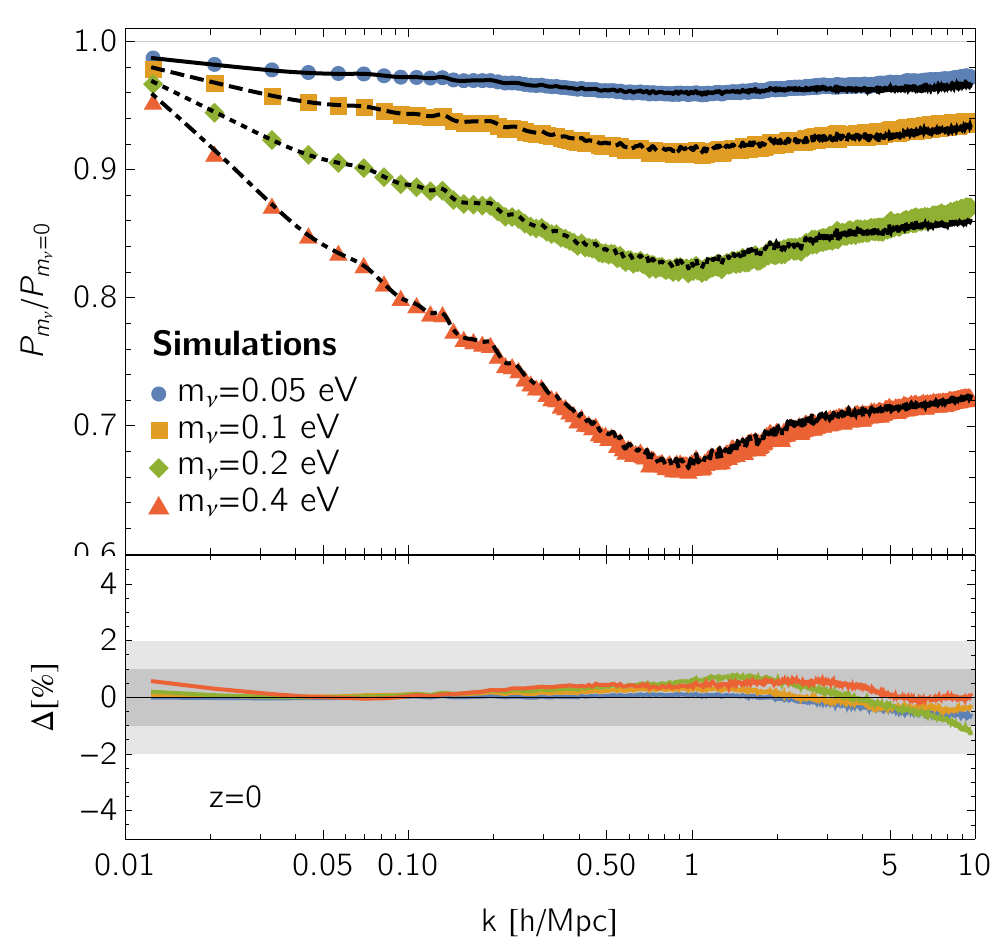}
\quad
\includegraphics[width=\columnwidth]{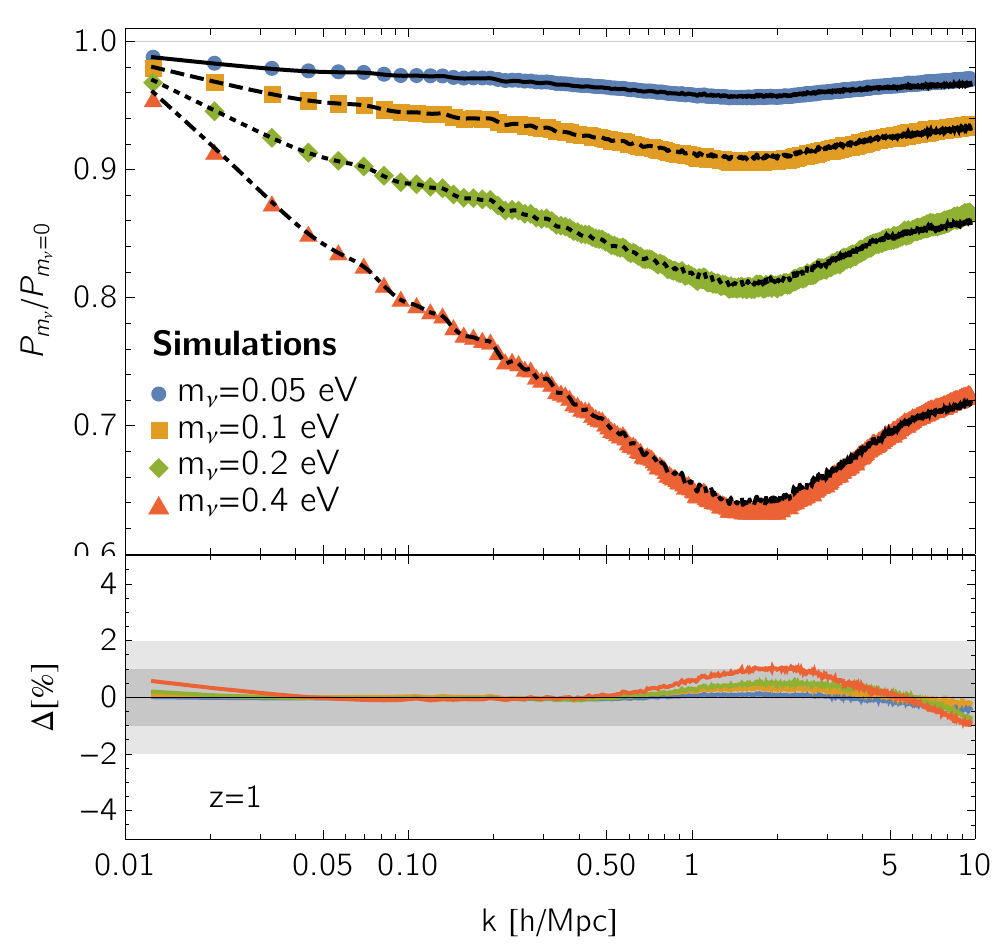}
\end{center}
\caption{
Total matter power spectrum ratios of the massive to the massless neutrino cosmologies at $z=0$ (left) and $z=1$ (right). The data points show the results of the $L_{\rm box} = 500 \,  \Mpch$ simulations described in Sec.~\ref{sec:sims}, and the black lines correspond to the halo model reaction predictions, $P^{\rm (m)} = \mathcal{R} \times P^{\rm pseudo}$, where $P^{\rm pseudo}$ is taken from flat $\Lambda$CDM dark matter-only simulations with \emph{pseudo} initial conditions, and the halo model reactions are computed assuming the ~\citet{Despali:2016} and~\citet{Bullock:2001} fits for the halo mass functions and $c$-$M$ relations, respectively. The lower panels illustrate the excellent performance of our method, which matches the simulations at percent level for all $k \lesssim 10 \, \hMpc$ (solid lines). 
}
\label{fig:pk_ratios}
\end{figure*}
Halo catalogues for each simulation are generated using a spherical overdensity algorithm based on the method described in \citet{HarnoisDeraps:2013}. Briefly, the first stage of this process is to identify halo candidates as peaks in the dark matter density field. This is achieved by interpolating dark matter particles onto a uniform mesh with cell width $81 \, h^{-1}{\rm kpc}$ and denoting candidates as local maxima in the density field. We then refine the density interpolation in the local region of each candidate using a mesh of width $16 \, h^{-1}{\rm kpc}$ and identify a centre as the location of maximum density. The halo radius is defined by building spherical shells around the centre until the enclosed density reaches the cosmology- and redshift-dependent virial density, $\Delta_{\mathrm{vir}}$, derived from the spherical collapse and virial theorem. The density profile for each halo is stored using 20 logarithmically-spaced bins that reach out to $2 \, \Mpch$. We compute a concentration for each halo by performing a least-squares fit to an NFW density profile. When doing so, we discard all radial bins smaller than twice the gravitational softening length and larger than the virial radius.


\section{Results}\label{sec:results}
\subsection{$P^{\mathrm(m)}$ from the standard halo abundance and concentration fits}
\begin{figure*}
\begin{center}
\includegraphics[width=\columnwidth]{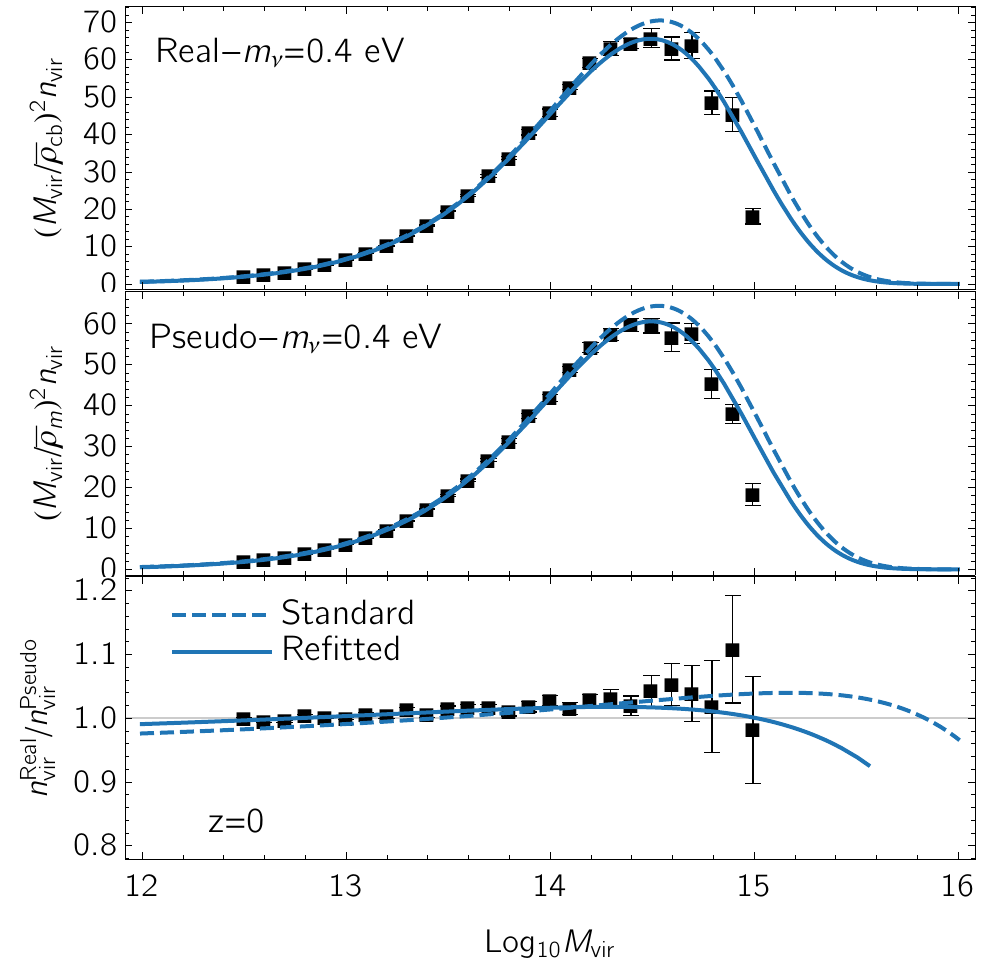}
\quad
\includegraphics[width=0.965\columnwidth]{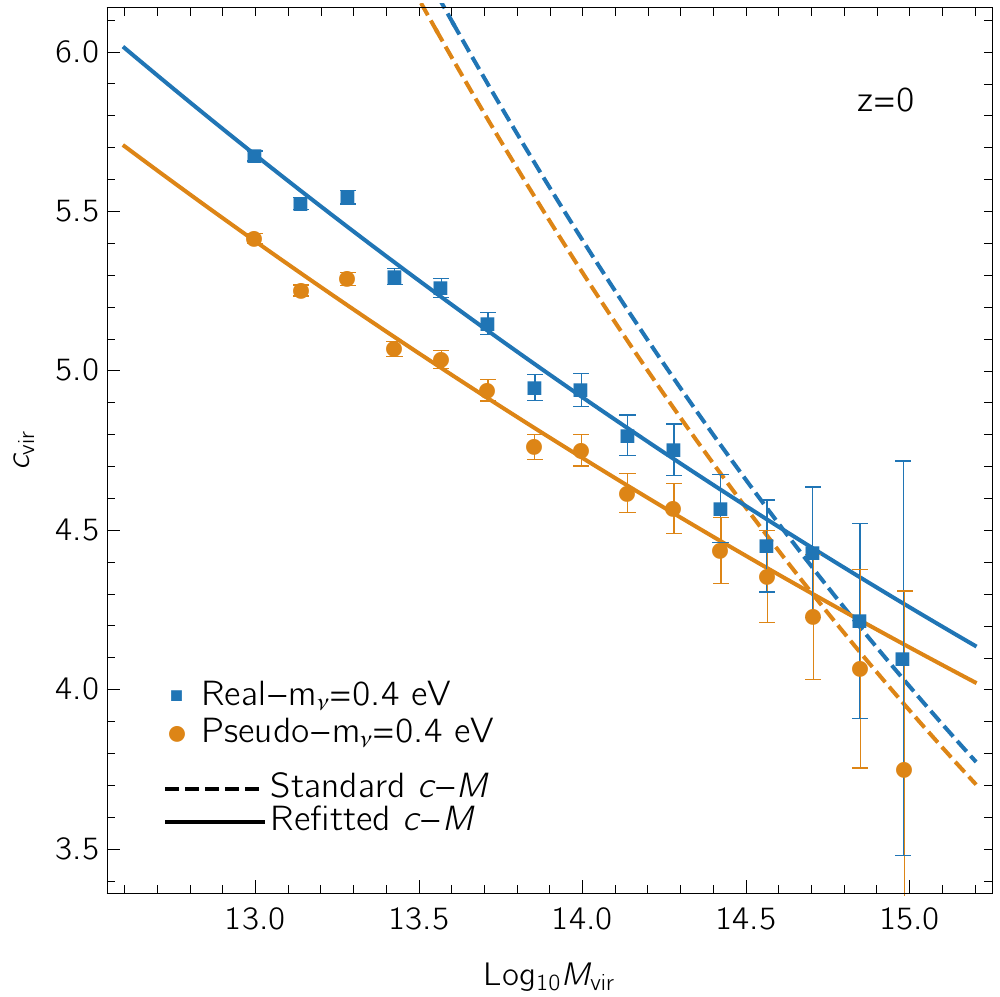}
\end{center}
\caption{Halo properties extracted from the $z=0$ snapshots of the large-volume simulations ($L_{\rm box} = 1000 \, \Mpch$). {\it Left:} the abundance of dark matter halos for the real (top panel) and pseudo (middle panel) cosmologies with $m_\nu = 0.4$ eV, both adjusted with prefactors such as to match the large-scale limit of the corresponding one-halo integrands (Eq.~\ref{eq:P1h}). The lower panel shows the \emph{real}-to-\emph{pseudo} halo mass function ratio, a quantity controlling the two-to-one-halo transition of the \emph{halo model reaction}. The data points and error bars represent the means and Jackknife uncertainties obtained by splitting the simulation boxes in octants. Halo masses are binned in logarithmic bins of size $\Delta\log_{10}M = 0.1$. We only use halos with more than 1000 particles and discard mass bins with fewer than 5 halos per sub-volume. The blue lines represent the Sheth-Tormen semi-analytical predictions with halo mass function parameters either taken from~\citet{Despali:2016} (dashed) or re-calibrated to fit individually our \emph{real} and \emph{pseudo} simulations (solid). {\it Right:} virial concentration-mass relation for the \emph{real} (blue) and \emph{pseudo} (orange) cosmologies with $m_\nu = 0.4$ eV. The coloured lines are power law approximations with parameter values taken from~\citet{Bullock:2001} (dashed) or fitted to our simulations (solid). Symbols denote measurements from the simulations with central values corresponding to the mass-weighted mean concentration of the halos within each mass bin, and error bars only account for the Poisson noise. 
In addition, we only keep halos with more than 3000 particles to minimise profile fitting errors. 
}
\label{fig:sim_virial}
\end{figure*}

We begin by presenting the performance of the \emph{halo model reactions} against our suite of small-volume simulations. For this comparison our reaction predictions (Eq.~\ref{eq:react}) are based on the standard values of the parameters entering the halo mass function~\citep{Despali:2016} and $c$-$M$ relation~\citep{Bullock:2001}, which we apply to both the \emph{real} and \emph{pseudo} massive neutrino cosmologies. The upper panels of Fig.~\ref{fig:pk_ratios} shows the the impact of massive neutrinos on the non-linear total matter power spectrum for the range of neutrino masses relevant for the next generation of cosmological surveys~\citep{Coulton:2019}. The lower panels display the relative deviation of our predictions (see Eq.~\ref{eq:tot_matter_pred}) from the full massive neutrino simulations, which is $\lesssim 1\%$ over the entire range of scales analysed and at both redshifts considered. This highly accurate result follows from the good agreement between the predicted \emph{real}-to-\emph{pseudo} halo mass function ratio and the simulations, which we show in the lower-left panel of Fig.~\ref{fig:sim_virial} for the largest neutrino mass in our study. \citet{Cataneo:2019} noticed that this quantity is directly related to the accuracy of the \emph{reaction} across the transition to the non-linear regime. In fact, although the \emph{real} and \emph{pseudo} standard halo mass functions are a poor fit for halo masses $M \gtrsim 10^{14.5} \, h^{-1}M_\odot$ when taken individually (Fig.~\ref{fig:sim_virial}, upper- and middle-left panel), the predicted ratio remains within $\sim 2\%$ of the simulation measurements, thus corroborating the original findings of~\citet{Cataneo:2019}. On the other hand, halo concentrations become relevant deep in the non-linear regime, and the right panel of Fig.~\ref{fig:sim_virial} illustrates that despite the large absolute inaccuracies of the standard fits, once again the \emph{real}-to-\emph{pseudo} ratio is not too dissimilar from that of the simulations. This fact enables the excellent performance of the \emph{halo model reactions} on scales $k \gtrsim 1 \, \hMpc$. 

\begin{figure}
\begin{center}
\includegraphics[width=\columnwidth]{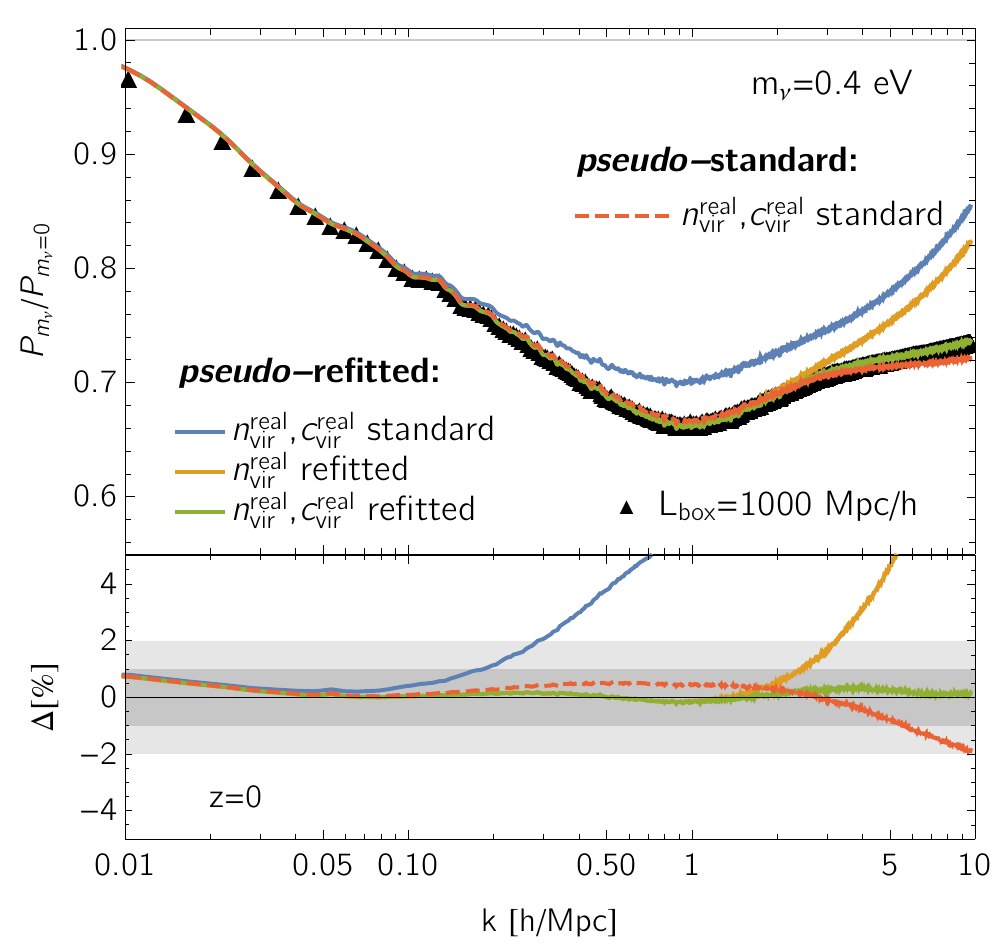}
\end{center}
\caption{Present-day total matter power spectrum ratio of the massive neutrino cosmology with $m_\nu = 0.4$ eV relative to the massless neutrino case. Symbols correspond to the measurements from the large-volume simulations. Solid lines are the \emph{halo model reaction} predictions adopting the refitted halo mass functions and $c$-$M$ relations shown in Fig.~\ref{fig:sim_virial} for the \emph{pseudo} cosmology, while the \emph{real} quantities use either the standard or refitted parameters. For comparison, we also show the predictions computed using the standard fits for both the \emph{pseudo} and the \emph{real} halo properties (dashed line). For all cases, our predictions use the non-linear matter power spectrum of the large-volume \emph{pseudo} simulation. The lower panel shows that once the \emph{pseudo} halo properties are calibrated to the simulations, the \emph{reaction} enables an accurate one-to-one mapping between the \emph{real} halo properties and the power spectrum, thus out-performing the traditional halo model calculations. Differences on small scales for the predictions based on the full standard fits (dashed line) compared to those in Fig.~\ref{fig:pk_ratios} are due to different halo concentrations in the small- and large-volume \emph{real} massive neutrino simulations. 
}
\label{fig:MV}
\end{figure}
\subsection{The effect of halo properties measured in simulations}\label{sec:fits}
It is currently unclear how accurately the non-linear matter power spectrum can be predicted given just mean halo properties such as their abundance and density profiles.  For the standard halo model, it is well known that this approach fails due to large inaccuracies on quasi-linear scales of the absolute power spectrum~\citep[see, e.g.,][]{Giocoli:2010,Massara:2014}. The \emph{halo model reactions}, however, are fractional quantities, and as such better suited to absorb the errors incurred separately by the \emph{real} and \emph{pseudo} halo model predictions.  To quantify the accuracy of this approach, we fit the Sheth-Tormen mass function and $c$-$M$ relations to our large volume $m_\nu=0.4$ eV simulations, obtaining $\{A^\mathrm{real},q^\mathrm{real},p^\mathrm{real}\} = \{0.3152,0.8423,0\}$, $\{A^\mathrm{pseudo},q^\mathrm{pseudo},p^\mathrm{pseudo}\} = \{0.3097,0.8313,0\}$, $\{c_0^\mathrm{real},\alpha^\mathrm{real}\} = \{6.3,0.062\}$, $\{c_0^\mathrm{pseudo},\alpha^\mathrm{pseudo}\} = \{6,0.058\}$. We show these fits as solid lines in Fig.~\ref{fig:sim_virial}. 

To estimate the relative importance of the mean halo properties for the accuracy of the predicted non-linear power spectrum, in Fig.~\ref{fig:MV} we fix the \emph{pseudo} halo mass function and $c$-$M$ parameters to their refitted values while varying their \emph{real} counterparts. When the parameters entering the halo mass function and $c$-$M$ relation are all set to their standard values (blue line), our predictions experience deviations as large as $\sim 10\%$ for $k \gtrsim 0.1 \, \hMpc$. The match to the simulations improves substantially on scales $0.1 \lesssim k \lesssim 1 \, \hMpc$ by including information on the halo mass function (orange line). If we further add our knowledge of the \emph{real} halo concentrations, the agreement with the simulations reaches sub-percent level down to the smallest scales modelled in this study. These results confirm that the \emph{halo model reactions} can produce even higher-quality predictions when supplied with accurate halo properties and \emph{pseudo} non-linear power spectra\footnote{As pointed out earlier in the text, the \emph{reactions} are fractional quantities, that is, as long as the same halo finder and halo concentration algorithm are used for the \emph{real} and \emph{pseudo} cosmologies, the refitted halo-model predictions will match the simulations very well. In the future we will be interested in calibrating the \emph{pseudo} halo properties with the end goal of building emulators. At that stage, the level of convergence in the output of more sophisticated halo finders~\citep[e.g.,][]{Behroozi:2013,Elahi:2019} will be an important indicator of the absolute accuracy attainable by the \emph{reaction} framework.}. For comparison, we also show the calculation based on the standard fits for both the \emph{pseudo} and the \emph{real} halo properties (dashed line). Differences on scales $k \gtrsim 5 \, \hMpc$ compared to the same prediction in Fig.~\ref{fig:pk_ratios} are primarily sourced by changes to the concentrations of small halos between the small- and large-volume simulations of the \emph{real} massive neutrino cosmology, which in turn depend on the different $N_\nu/N_{\rm cb}$ particle number ratio used for these two runs (see Sec.~\ref{sec:sims}).

\subsection{Comparison to \sc{halofit}}\label{sec:halofit}
We shall now assess the validity of the \emph{halo model reactions} for alternative implementations of the gravitational force~\citep[e.g.][]{Springel:2005,Habib:2016} and of massive neutrinos~\citep[e.g.][]{Banerjee:2016,Ali-Haimoud:2018} in $N$-body codes. Ideally, we would carry out this test using the simulation outputs of codes other than {\sc cubep$^3$m}~\citep[e.g.,][]{Castorina:2015,Liu:2018}. However, publicly available snapshots do not include runs for the \emph{pseudo} cosmologies, which means we must resort to our simulations for these cases. Given that the clustering of matter generated by different codes can vary considerably even for dark matter-only simulations~\citep{Schneider:2016,Garrison:2019}, this choice could bias our conclusions in the highly non-linear regime. Instead, we use {\sc halofit} to compute the non-linear matter power spectrum, employing the \citet{Takahashi:2012} calibration for the {\it pseudo} and the massless $\Lambda$CDM cases, and the \citet{Bird:2012} prescription for the massive neutrino cosmologies; these two fitting functions are calibrated to the output of {\sc gadget-2} and {\sc gadget-3} codes~\citep{Springel:2001,Springel:2005}, respectively.
Moreover, for this comparison we use the standard halo mass function and $c$-$M$ relation parameters listed in Sec.~\ref{sec:reactions}, i.e. without refitting to the {\sc cubep$^3$m} simulations. We find that our \emph{reaction}-based predictions for the total matter power spectrum of the massive neutrino cosmologies deviate no more than 3\% from the {\sc halofit} outputs. Such departures are comparable to, or smaller than, the typical {\sc halofit} inaccuracies~\citep[see, e.g.,][]{Knabenhans:2019,Smith:2019}, which suggests our method can also satisfactorily reproduce the results of other $N$-body codes provided that the baseline \emph{pseudo} power spectrum is obtained from simulations run with the same code and initial random phases of their \emph{real} massive neutrino counterparts.

\section{Discussion}\label{sec:discussion}
In this paper we incorporated in the \emph{halo model reaction} framework of~\citet{Cataneo:2019} an effective analytical strategy to accurately describe the non-linear effects induced by massive neutrinos on the total matter power spectrum. Our approach draws from the \emph{cold dark matter prescription} adopted in~\cite{Massara:2014}, with the notable difference that here we treated the clustering of massive neutrinos as purely linear, and worked with \emph{pseudo} rather than the standard massless neutrino cosmology as baseline in our halo model power spectrum ratios. In contrast to modified gravity cosmologies~\citep{Cataneo:2019}, we found that the inclusion of high-order perturbative corrections to the two-halo contributions in the \emph{reaction} was unnecessary. 

We studied the interdependency between halo properties and matter power spectrum \emph{reactions}, and conclusively showed that accurate knowledge of the mean halo abundances and concentrations (both central in cluster cosmology studies) leads to exquisite predictions for the \emph{halo model reactions}. Together with the fast emulation method to compute the \emph{pseudo} non-linear matter power spectrum presented in~\citet{Giblin:2019}, the tight connection between halo mass function and matter power spectrum in our approach enables, for instance, the simultaneous analysis of cluster number counts and cosmic shear data in a novel, self-consistent way. In a future work, we will merge in a single \emph{reaction} function both massive neutrino and dark energy/modified gravity cosmologies, which will enable us to predict the combined effects of these extensions on the matter power spectrum in a regime so far only accessible to specially modified $N$-body simulations~\citep{Baldi:2014,Giocoli:2018,Wright:2019}.

Finally, poorly understood baryonic processes impact the distribution of matter on scales $k \gtrsim 1 \, \hMpc$, thus limiting our ability to correctly model the power spectrum deep in the non-linear regime~\citep[see][for a review]{Chisari:2019}. It was showed that it is possible to account for these additional effects within the halo model~\citep{Semboloni:2011,Semboloni:2013,Fedeli:2014,Mohammed:2014,Mead:2015,Schneider:2019,Debackere:2019}, and we leave the implementation of baryonic feedback in the \emph{halo model reactions} to future investigation.

\section*{Acknowledgements}
MC thanks A. Mead for useful conversations in the early stages of this work.
MC, JHD and CH acknowledge support from the European Research Council under grant number 647112.
CH acknowledges support from the Max Planck Society and the Alexander von Humboldt Foundation in the framework of the Max Planck-Humboldt Research Award endowed by the Federal Ministry of Education and Research.
Work at Argonne National Laboratory was supported under U.S. Department of Energy contract DE-AC02-06CH11357.
The authors are grateful to Ue-Li Pen for his assistance with computing
resources.
Computations were performed on the BGQ and Niagara supercomputers at the SciNet HPC Consortium \citep{Loken:2010,Ponce:2019}. SciNet is funded by: the Canada Foundation for Innovation; the Government of Ontario; Ontario Research Fund - Research Excellence; and the University of Toronto.





\bibliographystyle{mnras}
\bibliography{references} 

\begin{thebibliography}{}
\makeatletter
\relax
\def\mn@urlcharsother{\let\do\@makeother \do\$\do\&\do\#\do\^\do\_\do\%\do\~}
\def\mn@doi{\begingroup\mn@urlcharsother \@ifnextchar [ {\mn@doi@}
  {\mn@doi@[]}}
\def\mn@doi@[#1]#2{\def\@tempa{#1}\ifx\@tempa\@empty \href
  {http://dx.doi.org/#2} {doi:#2}\else \href {http://dx.doi.org/#2} {#1}\fi
  \endgroup}
\def\mn@eprint#1#2{\mn@eprint@#1:#2::\@nil}
\def\mn@eprint@arXiv#1{\href {http://arxiv.org/abs/#1} {{\tt arXiv:#1}}}
\def\mn@eprint@dblp#1{\href {http://dblp.uni-trier.de/rec/bibtex/#1.xml}
  {dblp:#1}}
\def\mn@eprint@#1:#2:#3:#4\@nil{\def\@tempa {#1}\def\@tempb {#2}\def\@tempc
  {#3}\ifx \@tempc \@empty \let \@tempc \@tempb \let \@tempb \@tempa \fi \ifx
  \@tempb \@empty \def\@tempb {arXiv}\fi \@ifundefined
  {mn@eprint@\@tempb}{\@tempb:\@tempc}{\expandafter \expandafter \csname
  mn@eprint@\@tempb\endcsname \expandafter{\@tempc}}}

\bibitem[\protect\citeauthoryear{{Abazajian}, {Switzer}, {Dodelson}, {Heitmann}
   \& {Habib}}{{Abazajian} et~al.}{2005}]{Abazajian:2005}
{Abazajian} K.,  {Switzer} E.~R.,  {Dodelson} S.,  {Heitmann} K.,   {Habib} S.,
   2005, \mn@doi [\prd] {10.1103/PhysRevD.71.043507}, \href
  {https://ui.adsabs.harvard.edu/abs/2005PhRvD..71d3507A} {71, 043507}

\bibitem[\protect\citeauthoryear{{Agarwal} \& {Feldman}}{{Agarwal} \&
  {Feldman}}{2011}]{Agarwal:2011}
{Agarwal} S.,  {Feldman} H.~A.,  2011, \mn@doi [\mnras]
  {10.1111/j.1365-2966.2010.17546.x}, \href
  {https://ui.adsabs.harvard.edu/abs/2011MNRAS.410.1647A} {410, 1647}

\bibitem[\protect\citeauthoryear{{Ahmed} et~al.,}{{Ahmed}
  et~al.}{2004}]{Ahmed:2004}
{Ahmed} S.~N.,  et~al., 2004, \mn@doi [\prl] {10.1103/PhysRevLett.92.181301},
  \href {https://ui.adsabs.harvard.edu/abs/2004PhRvL..92r1301A} {92, 181301}

\bibitem[\protect\citeauthoryear{{Ali-Ha{\"\i}moud} \&
  {Bird}}{{Ali-Ha{\"\i}moud} \& {Bird}}{2013}]{Ali-Haimoud:2013}
{Ali-Ha{\"\i}moud} Y.,  {Bird} S.,  2013, \mn@doi [\mnras]
  {10.1093/mnras/sts286}, \href
  {https://ui.adsabs.harvard.edu/abs/2013MNRAS.428.3375A} {428, 3375}

\bibitem[\protect\citeauthoryear{{Baldi}, {Villaescusa-Navarro}, {Viel},
  {Puchwein}, {Springel}  \& {Moscardini}}{{Baldi} et~al.}{2014}]{Baldi:2014}
{Baldi} M.,  {Villaescusa-Navarro} F.,  {Viel} M.,  {Puchwein} E.,  {Springel}
  V.,   {Moscardini} L.,  2014, \mn@doi [\mnras] {10.1093/mnras/stu259}, \href
  {https://ui.adsabs.harvard.edu/abs/2014MNRAS.440...75B} {440, 75}

\bibitem[\protect\citeauthoryear{{Banerjee} \& {Dalal}}{{Banerjee} \&
  {Dalal}}{2016}]{Banerjee:2016}
{Banerjee} A.,  {Dalal} N.,  2016, \mn@doi [\jcap]
  {10.1088/1475-7516/2016/11/015}, \href
  {https://ui.adsabs.harvard.edu/abs/2016JCAP...11..015B} {2016, 015}

\bibitem[\protect\citeauthoryear{{Behroozi}, {Wechsler}  \& {Wu}}{{Behroozi}
  et~al.}{2013}]{Behroozi:2013}
{Behroozi} P.~S.,  {Wechsler} R.~H.,   {Wu} H.-Y.,  2013, \mn@doi [\apj]
  {10.1088/0004-637X/762/2/109}, \href
  {http://adsabs.harvard.edu/abs/2013ApJ...762..109B} {762, 109}

\bibitem[\protect\citeauthoryear{{Bird}, {Viel}  \& {Haehnelt}}{{Bird}
  et~al.}{2012}]{Bird:2012}
{Bird} S.,  {Viel} M.,   {Haehnelt} M.~G.,  2012, \mn@doi [\mnras]
  {10.1111/j.1365-2966.2011.20222.x}, \href
  {https://ui.adsabs.harvard.edu/abs/2012MNRAS.420.2551B} {420, 2551}

\bibitem[\protect\citeauthoryear{{Bird}, {Ali-Ha{\"\i}moud}, {Feng}  \&
  {Liu}}{{Bird} et~al.}{2018}]{Ali-Haimoud:2018}
{Bird} S.,  {Ali-Ha{\"\i}moud} Y.,  {Feng} Y.,   {Liu} J.,  2018, \mn@doi
  [\mnras] {10.1093/mnras/sty2376}, \href
  {https://ui.adsabs.harvard.edu/abs/2018MNRAS.481.1486B} {481, 1486}

\bibitem[\protect\citeauthoryear{{Blas}, {Garny}, {Konstandin}  \&
  {Lesgourgues}}{{Blas} et~al.}{2014}]{Blas:2014}
{Blas} D.,  {Garny} M.,  {Konstandin} T.,   {Lesgourgues} J.,  2014, \mn@doi
  [\jcap] {10.1088/1475-7516/2014/11/039}, \href
  {https://ui.adsabs.harvard.edu/abs/2014JCAP...11..039B} {2014, 039}

\bibitem[\protect\citeauthoryear{{Bond}, {Efstathiou}  \& {Silk}}{{Bond}
  et~al.}{1980}]{Bond:1980}
{Bond} J.~R.,  {Efstathiou} G.,   {Silk} J.,  1980, \mn@doi [\prl]
  {10.1103/PhysRevLett.45.1980}, \href
  {https://ui.adsabs.harvard.edu/abs/1980PhRvL..45.1980B} {45, 1980}

\bibitem[\protect\citeauthoryear{{Bullock}, {Kolatt}, {Sigad}, {Somerville},
  {Kravtsov}, {Klypin}, {Primack}  \& {Dekel}}{{Bullock}
  et~al.}{2001}]{Bullock:2001}
{Bullock} J.~S.,  {Kolatt} T.~S.,  {Sigad} Y.,  {Somerville} R.~S.,  {Kravtsov}
  A.~V.,  {Klypin} A.~A.,  {Primack} J.~R.,   {Dekel} A.,  2001, \mn@doi
  [\mnras] {10.1046/j.1365-8711.2001.04068.x}, \href
  {http://adsabs.harvard.edu/abs/2001MNRAS.321..559B} {321, 559}

\bibitem[\protect\citeauthoryear{{Castorina}, {Carbone}, {Bel}, {Sefusatti}  \&
  {Dolag}}{{Castorina} et~al.}{2015}]{Castorina:2015}
{Castorina} E.,  {Carbone} C.,  {Bel} J.,  {Sefusatti} E.,   {Dolag} K.,  2015,
  \mn@doi [\jcap] {10.1088/1475-7516/2015/07/043}, \href
  {https://ui.adsabs.harvard.edu/abs/2015JCAP...07..043C} {2015, 043}

\bibitem[\protect\citeauthoryear{{Cataneo}, {Lombriser}, {Heymans}, {Mead},
  {Barreira}, {Bose}  \& {Li}}{{Cataneo} et~al.}{2019}]{Cataneo:2019}
{Cataneo} M.,  {Lombriser} L.,  {Heymans} C.,  {Mead} A.~J.,  {Barreira} A.,
  {Bose} S.,   {Li} B.,  2019, \mn@doi [\mnras] {10.1093/mnras/stz1836}, \href
  {https://ui.adsabs.harvard.edu/abs/2019MNRAS.tmp.1778C} {p.~1778}

\bibitem[\protect\citeauthoryear{{Chisari} et~al.,}{{Chisari}
  et~al.}{2019}]{Chisari:2019}
{Chisari} N.~E.,  et~al., 2019, \mn@doi [The Open Journal of Astrophysics]
  {10.21105/astro.1905.06082}, \href
  {https://ui.adsabs.harvard.edu/abs/2019OJAp....2.....C} {2, 4}

\bibitem[\protect\citeauthoryear{{Cooray} \& {Sheth}}{{Cooray} \&
  {Sheth}}{2002}]{Cooray:2002}
{Cooray} A.,  {Sheth} R.,  2002, \mn@doi [\physrep]
  {10.1016/S0370-1573(02)00276-4}, \href
  {http://adsabs.harvard.edu/abs/2002PhR...372....1C} {372, 1}

\bibitem[\protect\citeauthoryear{{Costanzi}, {Villaescusa-Navarro}, {Viel},
  {Xia}, {Borgani}, {Castorina}  \& {Sefusatti}}{{Costanzi}
  et~al.}{2013}]{Costanzi:2013}
{Costanzi} M.,  {Villaescusa-Navarro} F.,  {Viel} M.,  {Xia} J.-Q.,  {Borgani}
  S.,  {Castorina} E.,   {Sefusatti} E.,  2013, \mn@doi [\jcap]
  {10.1088/1475-7516/2013/12/012}, \href
  {https://ui.adsabs.harvard.edu/abs/2013JCAP...12..012C} {2013, 012}

\bibitem[\protect\citeauthoryear{{Coulton}, {Liu}, {Madhavacheril}, {B{\"o}hm}
  \& {Spergel}}{{Coulton} et~al.}{2019}]{Coulton:2019}
{Coulton} W.~R.,  {Liu} J.,  {Madhavacheril} M.~S.,  {B{\"o}hm} V.,   {Spergel}
  D.~N.,  2019, \mn@doi [\jcap] {10.1088/1475-7516/2019/05/043}, \href
  {https://ui.adsabs.harvard.edu/abs/2019JCAP...05..043C} {2019, 043}

\bibitem[\protect\citeauthoryear{{Debackere}, {Schaye}  \&
  {Hoekstra}}{{Debackere} et~al.}{2019}]{Debackere:2019}
{Debackere} S. N.~B.,  {Schaye} J.,   {Hoekstra} H.,  2019, arXiv e-prints,
  \href {https://ui.adsabs.harvard.edu/abs/2019arXiv190805765D} {p.
  arXiv:1908.05765}

\bibitem[\protect\citeauthoryear{{Despali}, {Giocoli}, {Angulo}, {Tormen},
  {Sheth}, {Baso}  \& {Moscardini}}{{Despali} et~al.}{2016}]{Despali:2016}
{Despali} G.,  {Giocoli} C.,  {Angulo} R.~E.,  {Tormen} G.,  {Sheth} R.~K.,
  {Baso} G.,   {Moscardini} L.,  2016, \mn@doi [\mnras]
  {10.1093/mnras/stv2842}, \href
  {https://ui.adsabs.harvard.edu/abs/2016MNRAS.456.2486D} {456, 2486}

\bibitem[\protect\citeauthoryear{{Elahi}, {Ca{\~n}as}, {Poulton}, {Tobar},
  {Willis}, {Lagos}, {Power}  \& {Robotham}}{{Elahi} et~al.}{2019}]{Elahi:2019}
{Elahi} P.~J.,  {Ca{\~n}as} R.,  {Poulton} R. J.~J.,  {Tobar} R.~J.,  {Willis}
  J.~S.,  {Lagos} C. d.~P.,  {Power} C.,   {Robotham} A. S.~G.,  2019, \mn@doi
  [\pasa] {10.1017/pasa.2019.12}, \href
  {https://ui.adsabs.harvard.edu/abs/2019PASA...36...21E} {36, e021}

\bibitem[\protect\citeauthoryear{{Emberson} et~al.,}{{Emberson}
  et~al.}{2017}]{Emberson:2017}
{Emberson} J.~D.,  et~al., 2017, \mn@doi [Research in Astronomy and
  Astrophysics] {10.1088/1674-4527/17/8/85}, \href
  {https://ui.adsabs.harvard.edu/abs/2017RAA....17...85E} {17, 085}

\bibitem[\protect\citeauthoryear{{Fedeli}}{{Fedeli}}{2014}]{Fedeli:2014}
{Fedeli} C.,  2014, \mn@doi [\jcap] {10.1088/1475-7516/2014/04/028}, \href
  {https://ui.adsabs.harvard.edu/abs/2014JCAP...04..028F} {2014, 028}

\bibitem[\protect\citeauthoryear{{Fukuda} et~al.,}{{Fukuda}
  et~al.}{1998}]{Fukuda:1998}
{Fukuda} Y.,  et~al., 1998, \mn@doi [\prl] {10.1103/PhysRevLett.81.1158}, \href
  {https://ui.adsabs.harvard.edu/abs/1998PhRvL..81.1158F} {81, 1158}

\bibitem[\protect\citeauthoryear{{Garrison}, {Eisenstein}  \&
  {Pinto}}{{Garrison} et~al.}{2019}]{Garrison:2019}
{Garrison} L.~H.,  {Eisenstein} D.~J.,   {Pinto} P.~A.,  2019, \mn@doi [\mnras]
  {10.1093/mnras/stz634}, \href
  {https://ui.adsabs.harvard.edu/abs/2019MNRAS.485.3370G} {485, 3370}

\bibitem[\protect\citeauthoryear{{Giblin}, {Cataneo}, {Moews}  \&
  {Heymans}}{{Giblin} et~al.}{2019}]{Giblin:2019}
{Giblin} B.,  {Cataneo} M.,  {Moews} B.,   {Heymans} C.,  2019, arXiv e-prints,
  \href {https://ui.adsabs.harvard.edu/abs/2019arXiv190602742G} {p.
  arXiv:1906.02742}

\bibitem[\protect\citeauthoryear{{Giocoli}, {Bartelmann}, {Sheth}  \&
  {Cacciato}}{{Giocoli} et~al.}{2010}]{Giocoli:2010}
{Giocoli} C.,  {Bartelmann} M.,  {Sheth} R.~K.,   {Cacciato} M.,  2010, \mn@doi
  [\mnras] {10.1111/j.1365-2966.2010.17108.x}, \href
  {http://adsabs.harvard.edu/abs/2010MNRAS.408..300G} {408, 300}

\bibitem[\protect\citeauthoryear{{Giocoli}, {Baldi}  \& {Moscardini}}{{Giocoli}
  et~al.}{2018}]{Giocoli:2018}
{Giocoli} C.,  {Baldi} M.,   {Moscardini} L.,  2018, \mn@doi [\mnras]
  {10.1093/mnras/sty2465}, \href
  {https://ui.adsabs.harvard.edu/abs/2018MNRAS.481.2813G} {481, 2813}

\bibitem[\protect\citeauthoryear{{Green} et~al.,}{{Green}
  et~al.}{2012}]{Green:2012}
{Green} J.,  et~al., 2012, arXiv e-prints, \href
  {https://ui.adsabs.harvard.edu/abs/2012arXiv1208.4012G} {p. arXiv:1208.4012}

\bibitem[\protect\citeauthoryear{{Habib} et~al.,}{{Habib}
  et~al.}{2016}]{Habib:2016}
{Habib} S.,  et~al., 2016, \mn@doi [\na] {10.1016/j.newast.2015.06.003}, \href
  {https://ui.adsabs.harvard.edu/abs/2016NewA...42...49H} {42, 49}

\bibitem[\protect\citeauthoryear{{Harnois-D{\'e}raps}, {Pen}, {Iliev}, {Merz},
  {Emberson}  \& {Desjacques}}{{Harnois-D{\'e}raps}
  et~al.}{2013}]{HarnoisDeraps:2013}
{Harnois-D{\'e}raps} J.,  {Pen} U.-L.,  {Iliev} I.~T.,  {Merz} H.,  {Emberson}
  J.~D.,   {Desjacques} V.,  2013, \mn@doi [\mnras] {10.1093/mnras/stt1591},
  \href {https://ui.adsabs.harvard.edu/abs/2013MNRAS.436..540H} {436, 540}

\bibitem[\protect\citeauthoryear{{Inman}}{{Inman}}{2017}]{Inman:2017b}
{Inman} D.,  2017, PhD thesis, University of Toronto (Canada)

\bibitem[\protect\citeauthoryear{{Inman}, {Emberson}, {Pen}, {Farchi}, {Yu}  \&
  {Harnois-D{\'e}raps}}{{Inman} et~al.}{2015}]{Inman:2015}
{Inman} D.,  {Emberson} J.~D.,  {Pen} U.-L.,  {Farchi} A.,  {Yu} H.-R.,
  {Harnois-D{\'e}raps} J.,  2015, \mn@doi [\prd] {10.1103/PhysRevD.92.023502},
  \href {https://ui.adsabs.harvard.edu/abs/2015PhRvD..92b3502I} {92, 023502}

\bibitem[\protect\citeauthoryear{{Knabenhans} et~al.,}{{Knabenhans}
  et~al.}{2019}]{Knabenhans:2019}
{Knabenhans} M.,  et~al., 2019, \mn@doi [\mnras] {10.1093/mnras/stz197}, \href
  {https://ui.adsabs.harvard.edu/abs/2019MNRAS.484.5509K} {484, 5509}

\bibitem[\protect\citeauthoryear{{LSST Dark Energy Science
  Collaboration}}{{LSST Dark Energy Science Collaboration}}{2012}]{LSST:2012}
{LSST Dark Energy Science Collaboration} 2012, preprint, \href
  {http://adsabs.harvard.edu/abs/2012arXiv1211.0310L} {} (\mn@eprint {arXiv}
  {1211.0310})

\bibitem[\protect\citeauthoryear{{Laureijs} et~al.,}{{Laureijs}
  et~al.}{2011}]{Laureijs:2011}
{Laureijs} R.,  et~al., 2011, preprint, \href
  {http://adsabs.harvard.edu/abs/2011arXiv1110.3193L} {} (\mn@eprint {arXiv}
  {1110.3193})

\bibitem[\protect\citeauthoryear{{Lawrence} et~al.,}{{Lawrence}
  et~al.}{2017}]{Lawrence:2017}
{Lawrence} E.,  et~al., 2017, \mn@doi [\apj] {10.3847/1538-4357/aa86a9}, \href
  {https://ui.adsabs.harvard.edu/abs/2017ApJ...847...50L} {847, 50}

\bibitem[\protect\citeauthoryear{{Lesgourgues} \& {Pastor}}{{Lesgourgues} \&
  {Pastor}}{2006}]{Lesgourgues:2006}
{Lesgourgues} J.,  {Pastor} S.,  2006, \mn@doi [\physrep]
  {10.1016/j.physrep.2006.04.001}, \href
  {http://adsabs.harvard.edu/abs/2006PhR...429..307L} {429, 307}

\bibitem[\protect\citeauthoryear{{Levi} et~al.,}{{Levi}
  et~al.}{2013}]{Levi:2013}
{Levi} M.,  et~al., 2013, preprint, \href
  {http://adsabs.harvard.edu/abs/2013arXiv1308.0847L} {} (\mn@eprint {arXiv}
  {1308.0847})

\bibitem[\protect\citeauthoryear{{Lewis}, {Challinor}  \& {Lasenby}}{{Lewis}
  et~al.}{2000}]{Lewis:2000}
{Lewis} A.,  {Challinor} A.,   {Lasenby} A.,  2000, \mn@doi [\apj]
  {10.1086/309179}, \href {http://adsabs.harvard.edu/abs/2000ApJ...538..473L}
  {538, 473}

\bibitem[\protect\citeauthoryear{{Liu}, {Bird}, {Zorrilla Matilla}, {Hill},
  {Haiman}, {Madhavacheril}, {Petri}  \& {Spergel}}{{Liu}
  et~al.}{2018}]{Liu:2018}
{Liu} J.,  {Bird} S.,  {Zorrilla Matilla} J.~M.,  {Hill} J.~C.,  {Haiman} Z.,
  {Madhavacheril} M.~S.,  {Petri} A.,   {Spergel} D.~N.,  2018, \mn@doi [\jcap]
  {10.1088/1475-7516/2018/03/049}, \href
  {https://ui.adsabs.harvard.edu/abs/2018JCAP...03..049L} {2018, 049}

\bibitem[\protect\citeauthoryear{{LoVerde}}{{LoVerde}}{2014}]{LoVerde:2014}
{LoVerde} M.,  2014, \mn@doi [\prd] {10.1103/PhysRevD.90.083518}, \href
  {https://ui.adsabs.harvard.edu/abs/2014PhRvD..90h3518L} {90, 083518}

\bibitem[\protect\citeauthoryear{{Loken} et~al.,}{{Loken}
  et~al.}{2010}]{Loken:2010}
{Loken} C.,  et~al., 2010, in Journal of Physics Conference Series. p. 012026,
  \mn@doi{10.1088/1742-6596/256/1/012026}

\bibitem[\protect\citeauthoryear{{Mangano}, {Miele}, {Pastor}, {Pinto},
  {Pisanti}  \& {Serpico}}{{Mangano} et~al.}{2005}]{Mangano:2005}
{Mangano} G.,  {Miele} G.,  {Pastor} S.,  {Pinto} T.,  {Pisanti} O.,
  {Serpico} P.~D.,  2005, \mn@doi [Nuclear Physics B]
  {10.1016/j.nuclphysb.2005.09.041}, \href
  {https://ui.adsabs.harvard.edu/abs/2005NuPhB.729..221M} {729, 221}

\bibitem[\protect\citeauthoryear{{Marulli}, {Carbone}, {Viel}, {Moscardini}  \&
  {Cimatti}}{{Marulli} et~al.}{2011}]{Marulli:2011}
{Marulli} F.,  {Carbone} C.,  {Viel} M.,  {Moscardini} L.,   {Cimatti} A.,
  2011, \mn@doi [\mnras] {10.1111/j.1365-2966.2011.19488.x}, \href
  {https://ui.adsabs.harvard.edu/abs/2011MNRAS.418..346M} {418, 346}

\bibitem[\protect\citeauthoryear{{Massara}, {Villaescusa-Navarro}  \&
  {Viel}}{{Massara} et~al.}{2014}]{Massara:2014}
{Massara} E.,  {Villaescusa-Navarro} F.,   {Viel} M.,  2014, \mn@doi [\jcap]
  {10.1088/1475-7516/2014/12/053}, \href
  {http://adsabs.harvard.edu/abs/2014JCAP...12..053M} {12, 053}

\bibitem[\protect\citeauthoryear{{McCarthy}, {Bird}, {Schaye},
  {Harnois-Deraps}, {Font}  \& {van Waerbeke}}{{McCarthy}
  et~al.}{2018}]{McCarthy:2018}
{McCarthy} I.~G.,  {Bird} S.,  {Schaye} J.,  {Harnois-Deraps} J.,  {Font}
  A.~S.,   {van Waerbeke} L.,  2018, \mn@doi [\mnras] {10.1093/mnras/sty377},
  \href {https://ui.adsabs.harvard.edu/abs/2018MNRAS.476.2999M} {476, 2999}

\bibitem[\protect\citeauthoryear{{Mead}}{{Mead}}{2017}]{Mead:2017}
{Mead} A.~J.,  2017, \mn@doi [\mnras] {10.1093/mnras/stw2312}, \href
  {http://adsabs.harvard.edu/abs/2017MNRAS.464.1282M} {464, 1282}

\bibitem[\protect\citeauthoryear{{Mead}, {Peacock}, {Heymans}, {Joudaki}  \&
  {Heavens}}{{Mead} et~al.}{2015}]{Mead:2015}
{Mead} A.~J.,  {Peacock} J.~A.,  {Heymans} C.,  {Joudaki} S.,   {Heavens}
  A.~F.,  2015, \mn@doi [\mnras] {10.1093/mnras/stv2036}, \href
  {http://adsabs.harvard.edu/abs/2015MNRAS.454.1958M} {454, 1958}

\bibitem[\protect\citeauthoryear{{Mead}, {Heymans}, {Lombriser}, {Peacock},
  {Steele}  \& {Winther}}{{Mead} et~al.}{2016}]{Mead:2016}
{Mead} A.~J.,  {Heymans} C.,  {Lombriser} L.,  {Peacock} J.~A.,  {Steele}
  O.~I.,   {Winther} H.~A.,  2016, \mn@doi [\mnras] {10.1093/mnras/stw681},
  \href {https://ui.adsabs.harvard.edu/abs/2016MNRAS.459.1468M} {459, 1468}

\bibitem[\protect\citeauthoryear{{Mohammed}, {Martizzi}, {Teyssier}  \&
  {Amara}}{{Mohammed} et~al.}{2014}]{Mohammed:2014}
{Mohammed} I.,  {Martizzi} D.,  {Teyssier} R.,   {Amara} A.,  2014, arXiv
  e-prints, \href {https://ui.adsabs.harvard.edu/abs/2014arXiv1410.6826M} {p.
  arXiv:1410.6826}

\bibitem[\protect\citeauthoryear{{Navarro}, {Frenk}  \& {White}}{{Navarro}
  et~al.}{1996}]{Navarro:1996}
{Navarro} J.~F.,  {Frenk} C.~S.,   {White} S.~D.~M.,  1996, \mn@doi [\apj]
  {10.1086/177173}, \href {http://adsabs.harvard.edu/abs/1996ApJ...462..563N}
  {462, 563}

\bibitem[\protect\citeauthoryear{{Ponce} et~al.,}{{Ponce}
  et~al.}{2019}]{Ponce:2019}
{Ponce} M.,  et~al., 2019, arXiv e-prints, \href
  {https://ui.adsabs.harvard.edu/abs/2019arXiv190713600P} {p. arXiv:1907.13600}

\bibitem[\protect\citeauthoryear{{Qian} \& {Vogel}}{{Qian} \&
  {Vogel}}{2015}]{Qian:2015}
{Qian} X.,  {Vogel} P.,  2015, \mn@doi [Progress in Particle and Nuclear
  Physics] {10.1016/j.ppnp.2015.05.002}, \href
  {https://ui.adsabs.harvard.edu/abs/2015PrPNP..83....1Q} {83, 1}

\bibitem[\protect\citeauthoryear{{Schmidt}}{{Schmidt}}{2016}]{Schmidt:2016}
{Schmidt} F.,  2016, \mn@doi [\prd] {10.1103/PhysRevD.93.063512}, \href
  {http://adsabs.harvard.edu/abs/2016PhRvD..93f3512S} {93, 063512}

\bibitem[\protect\citeauthoryear{{Schneider} et~al.,}{{Schneider}
  et~al.}{2016}]{Schneider:2016}
{Schneider} A.,  et~al., 2016, \mn@doi [\jcap] {10.1088/1475-7516/2016/04/047},
  \href {https://ui.adsabs.harvard.edu/abs/2016JCAP...04..047S} {2016, 047}

\bibitem[\protect\citeauthoryear{{Schneider}, {Teyssier}, {Stadel}, {Chisari},
  {Le Brun}, {Amara}  \& {Refregier}}{{Schneider}
  et~al.}{2019}]{Schneider:2019}
{Schneider} A.,  {Teyssier} R.,  {Stadel} J.,  {Chisari} N.~E.,  {Le Brun} A.
  M.~C.,  {Amara} A.,   {Refregier} A.,  2019, \mn@doi [\jcap]
  {10.1088/1475-7516/2019/03/020}, \href
  {https://ui.adsabs.harvard.edu/abs/2019JCAP...03..020S} {2019, 020}

\bibitem[\protect\citeauthoryear{{Semboloni}, {Hoekstra}, {Schaye}, {van
  Daalen}  \& {McCarthy}}{{Semboloni} et~al.}{2011}]{Semboloni:2011}
{Semboloni} E.,  {Hoekstra} H.,  {Schaye} J.,  {van Daalen} M.~P.,   {McCarthy}
  I.~G.,  2011, \mn@doi [\mnras] {10.1111/j.1365-2966.2011.19385.x}, \href
  {https://ui.adsabs.harvard.edu/abs/2011MNRAS.417.2020S} {417, 2020}

\bibitem[\protect\citeauthoryear{{Semboloni}, {Hoekstra}  \&
  {Schaye}}{{Semboloni} et~al.}{2013}]{Semboloni:2013}
{Semboloni} E.,  {Hoekstra} H.,   {Schaye} J.,  2013, \mn@doi [\mnras]
  {10.1093/mnras/stt1013}, \href
  {https://ui.adsabs.harvard.edu/abs/2013MNRAS.434..148S} {434, 148}

\bibitem[\protect\citeauthoryear{{Sheth} \& {Tormen}}{{Sheth} \&
  {Tormen}}{2002}]{Sheth:2002}
{Sheth} R.~K.,  {Tormen} G.,  2002, \mn@doi [\mnras]
  {10.1046/j.1365-8711.2002.04950.x}, \href
  {http://adsabs.harvard.edu/abs/2002MNRAS.329...61S} {329, 61}

\bibitem[\protect\citeauthoryear{{Smith} \& {Angulo}}{{Smith} \&
  {Angulo}}{2019}]{Smith:2019}
{Smith} R.~E.,  {Angulo} R.~E.,  2019, \mn@doi [\mnras] {10.1093/mnras/stz890},
  \href {https://ui.adsabs.harvard.edu/abs/2019MNRAS.486.1448S} {486, 1448}

\bibitem[\protect\citeauthoryear{{Springel}}{{Springel}}{2005}]{Springel:2005}
{Springel} V.,  2005, \mn@doi [\mnras] {10.1111/j.1365-2966.2005.09655.x},
  \href {https://ui.adsabs.harvard.edu/abs/2005MNRAS.364.1105S} {364, 1105}

\bibitem[\protect\citeauthoryear{{Springel}, {Yoshida}  \& {White}}{{Springel}
  et~al.}{2001}]{Springel:2001}
{Springel} V.,  {Yoshida} N.,   {White} S. D.~M.,  2001, \mn@doi [\na]
  {10.1016/S1384-1076(01)00042-2}, \href
  {https://ui.adsabs.harvard.edu/abs/2001NewA....6...79S} {6, 79}

\bibitem[\protect\citeauthoryear{{Takahashi}, {Sato}, {Nishimichi}, {Taruya}
  \& {Oguri}}{{Takahashi} et~al.}{2012}]{Takahashi:2012}
{Takahashi} R.,  {Sato} M.,  {Nishimichi} T.,  {Taruya} A.,   {Oguri} M.,
  2012, \mn@doi [\apj] {10.1088/0004-637X/761/2/152}, \href
  {https://ui.adsabs.harvard.edu/abs/2012ApJ...761..152T} {761, 152}

\bibitem[\protect\citeauthoryear{{Taylor}, {Kitching}  \& {McEwen}}{{Taylor}
  et~al.}{2018}]{Taylor:2018}
{Taylor} P.~L.,  {Kitching} T.~D.,   {McEwen} J.~D.,  2018, \mn@doi [\prd]
  {10.1103/PhysRevD.98.043532}, \href
  {http://adsabs.harvard.edu/abs/2018PhRvD..98d3532T} {98, 043532}

\bibitem[\protect\citeauthoryear{{Wright}, {Koyama}, {Winther}  \&
  {Zhao}}{{Wright} et~al.}{2019}]{Wright:2019}
{Wright} B.~S.,  {Koyama} K.,  {Winther} H.~A.,   {Zhao} G.-B.,  2019, \mn@doi
  [\jcap] {10.1088/1475-7516/2019/06/040}, \href
  {https://ui.adsabs.harvard.edu/abs/2019JCAP...06..040W} {2019, 040}

\bibitem[\protect\citeauthoryear{{Zel'Dovich}}{{Zel'Dovich}}{1970}]{Zeldovich:1970}
{Zel'Dovich} Y.~B.,  1970, \aap, \href
  {https://ui.adsabs.harvard.edu/abs/1970A&A.....5...84Z} {500, 13}

\makeatother
\end{thebibliography}




\appendix


\bsp	
\label{lastpage}
\end{document}